\documentclass[pra,twocolumn,showpacs,floatfix,amsfonts]{revtex4}
\usepackage{bm}
\usepackage{graphicx}  
\usepackage{here}                    
\usepackage{latexsym}                
\usepackage{amsmath}     
\usepackage{amsfonts}  
\usepackage{amssymb}  
\usepackage{dcolumn}

\newcommand{\bra}[1]{\langle #1 | \,}
\newcommand{\ket}[1]{\, | #1 \rangle}

\newcommand{\be}{\begin{equation}}
\newcommand{\ee}{\end{equation}}
\newcommand{\bea}{\begin{eqnarray}}
\newcommand{\eea}{\end{eqnarray}}

\def\unity{\openone}

\newcommand{\field}[1]{\mathbb{F}_{#1}}
\newcommand{\Er}[1]{\mathfrak{A}_{#1}}
\newcommand{\Hr}{\mathfrak{F}}

\newcommand{\PosInt}{\mathbb{N}}
\newcommand{\Int}{\mathbb{Z}}

\newcommand{\Real}{\mathbb{R}}
\begin{document}
\title{Error tolerance of two-basis quantum key-distribution protocols 
using qudits and two-way classical communication}
\author{Georgios M. Nikolopoulos}
\affiliation{Institut f\"ur Angewandte Physik, Technische 
Universit\"at Darmstadt, 64289 Darmstadt, Germany}
\author{Kedar S. Ranade}
\affiliation{Institut f\"ur Angewandte Physik, Technische 
Universit\"at Darmstadt, 64289 Darmstadt, Germany}
\author{Gernot Alber} 
\affiliation{Institut f\"ur Angewandte Physik, Technische 
Universit\"at Darmstadt, 64289 Darmstadt, Germany}

\date{\today}

\begin{abstract}
We investigate the error tolerance of quantum cryptographic 
protocols using $d$-level systems. In particular, we focus on 
prepare-and-measure schemes that use two mutually unbiased bases  
and a key-distillation procedure with two-way classical communication. 
For arbitrary quantum channels, we obtain a 
sufficient condition for secret-key distillation which, 
in the case of isotropic quantum channels, yields an analytic expression 
for the maximally tolerable error rate of the cryptographic protocols 
under consideration. 
The difference between the tolerable error rate and its 
theoretical upper bound tends slowly to zero for sufficiently large 
dimensions of the information carriers. 
\end{abstract}

\pacs{03.67.Dd, 03.67.Hk}

\maketitle

\section{Introduction}
 Provable entanglement has been shown to be a 
necessary precondition for secure quantum key-distribution (QKD) 
in the context of any protocol \cite{CLL,AG}. 
Recently \cite{NA}, we investigated the maximal average disturbance 
(error rate) up to which the two legitimate users (Alice and Bob) of a QKD protocol 
can prove the presence of quantum correlations in their sifted classical data. 
In particular, we focused on qudit-based QKD protocols using two Fourier-dual 
bases (to be referred to hereafter as $2d$-state protocols). 
Under the assumption of arbitrary joint (coherent) attacks we 
showed that the threshold disturbance for provable entanglement scales 
with the qudit-dimension $d$ as 
\bea
D_{\rm th}(d)=\frac{d-1}{2d}. 
\label{ThdistilConst}
\eea

This theoretical upper bound on tolerable error rates for $2d$-state protocols
is valid for arbitrary dimensions, provided that Alice and Bob 
focus on their sifted key and do not apply any collective measurements on 
their halves. 
Its implications are obvious for estimated disturbances above $D_{\rm th}$  
namely, Alice and Bob are not able to infer whether the correlations in their 
data have originated from an entangled state or not, and the protocol must 
be aborted. 
However, for detected disturbances below $D_{\rm th}$, the picture is 
incomplete. In particular, based on the above result we only know that the 
two honest parties can be confident that they share provable 
entanglement with high probability. Thus, the necessary precondition for 
secret-key distillation is satisfied for disturbances up to $D_{\rm th}$. 
Nevertheless, the details of a prepare-and-measure (P\&M) scheme which will be 
capable of attaining this theoretical bound are unknown. 
In fact, it is not at all clear whether such a P\&M scheme 
exists.    

So far, the highest tolerable error rates in the 
framework of P\&M QKD schemes have been reported for protocols 
using a two-way Gottesman-Lo-type procedure for key distillation \cite{GL}. 
This procedure was introduced  and improved in the 
context of the standard qubit-based $(d=2)$ QKD protocols \cite{GL,C-2}. 
It is based on local quantum operations and 
two-way classical communication (LOCC2) and is able to provide the 
two legitimate users with an unconditionally secure key up to high 
error rates. In particular for the standard $4$-state qubit-based protocol 
(BB84) the tolerable error rate is $20\%$ \cite{C-2,RA} which is well below the 
corresponding theoretical upper bound given by Eq. (\ref{ThdistilConst}), 
that is $25\%$. The natural question arises therefore whether this gap 
still persists for higher dimensions $(d>2)$ and, in particular, how it scales 
with the dimension $d$ of information carriers.

Recently, extending the Gottesman-Lo two-way key distillation (GL2KD)  
procedure to higher dimensions, Chau addressed this open question in the context of  
fully-symmetric qudit-based QKD schemes using all $(d+1)$ possible 
mutually unbiased bases \cite{C-d}. More precisely he showed that 
if $d$ is a prime power, the tolerable error-rate scales with dimension 
as $1-(3+\sqrt{5})/2d$, for $d\to\infty$. 

In this paper, our purpose is to analyze the error tolerance 
of $2d$-state QKD protocols using a GL2KD process. In contrast to the protocols 
considered in \cite{C-d}, the protocols considered here are not necessarily 
fully symmetric. 
In general, we have only one symmetry constraint i.e., the symmetry 
between the two Fourier-dual bases used in the protocol. 
Hence, the problem in its most general form is analytically solvable 
to some extent only. Specifically, we are able to derive a sufficient condition for 
secret-key distillation in which the number of open parameters  
scales quadratically with $d$. However, the derivation of an analytic expression 
for the tolerable error rate is possible under additional symmetry assumptions 
related to isotropic quantum channels. 
In this case, we find that the asymptotic $(d\to\infty)$ tolerable error-rate 
scales with dimension as $1/2-1/4\sqrt{d}$, and slowly approaches therefore 
its theoretical upper bound determined by Eq.(\ref{ThdistilConst}), that is $1/2$.

The organization of the paper follows the three phases of a  
typical P\&M QKD scheme. In Sec. \ref{intro-2}, for the sake of 
completeness we briefly summarize basic facts about the 
first two phases of a $2d$-state QKD protocol, i.e.,   
quantum state distribution and verification test. 
Subsequently, in Sec. \ref{secII} we focus on the key-distillation phase 
which is the main subject of this work. In particular 
we consider a GL2KD procedure. Our analysis is based on the 
entanglement-based version of the $2d$-state QKD protocol, whose 
reduction to a P\&M scheme is summarized at the end of the section. 
An analytic expression for the tolerable error-rate is derived in Sec. \ref{Sec-Iso} 
under the assumption of isotropic quantum channels.
Finally, we conclude with a short summary and outlook in Sec. \ref{secIV}.

\section{The first two stages of Two-basis QKD protocols}
\label{intro-2}
For the sake of simplicity, and without loss of generality, 
we will focus on prime dimensions only. Thus, throughout this work all the 
arithmetics are performed in the finite (Galois) field 
$\field{d}=\{0,1,\ldots,d-1\}$ \cite{ECC-book}. It has to be noted, however, 
that similar arguments hold if $d$ is a prime power but the formalism 
is more involved (e.g., see \cite{NA}).

In general, theoretical investigations of $d$-level quantum systems (qudits) 
are performed conveniently with the help of the generalized Pauli operators   
\bea
\Er{mn} := \sum_{l\in\field{d}} \Phi(l\cdot n)
\ket{l-m}\bra{l}\quad {\rm for}\, m,n\in\field{d},
\label{err2}
\eea
where  $\Phi(x)\equiv \exp(\frac{{\rm i}2\pi x}{d})$.
These $d^2$ operators form a faithful projective unitary representation 
of $\left (\Int\backslash d\Int\right )\times \left (\Int\backslash d\Int\right )$ 
and an error basis on the Hilbert space of a qudit $\mathbb{C}^d$ 
\cite{ErrorGroup}.

In a typical $2d$-state P\&M scheme, Alice and Bob use for their 
purposes two mutually unbiased bases. Following \cite{NA,BKBGC,AGS},
throughout this work we choose  the eigenbasis 
$\{\ket{\alpha} : \alpha\in\field{d}\}$ of $\Er{01}$ as the 
standard (computational) basis ${\cal B}_1$, while the second basis 
${\cal B}_2$ is the Fourier dual of the computational basis with the 
discrete Fourier transformation given by 
\[
\Hr := \frac{1}{\sqrt{d}}\sum_{i,j\in\field{d}}\Phi(i\cdot j) 
\ket{i}\bra{j}.
\] 
Hence, the indices $m$ and $n$ in Eq. (\ref{err2}), refer to dit-flip 
and phase errors in the standard basis ${\cal B}_1$, respectively. 
Moreover, $\Hr^\dag \Er{mn} \Hr = \Phi(-m\cdot n)\Er{nm}^*$ which 
indicates that dit-flip errors in the computational basis become phase 
errors in the complementary basis and vice-versa. 

In general, the first stage of a QKD protocol is the quantum state 
distribution stage which involves quantum state (signal) preparation 
and transmission via an insecure quantum channel.
The purpose of this phase is to establish correlations between 
Alice and Bob, which may also involve correlations 
with a third untrusted party (eavesdropper). 
As far as a typical $2d$-state P\&M scheme is concerned, 
this first stage proceeds 
as follows \cite{NA,C-d,BKBGC,AGS}. 
Alice sends to Bob a sequence of qudits each of which is 
randomly prepared in one of the $2d$ non-orthogonal basis-states 
($d$ states for each basis). Bob measures each received particle 
randomly in ${\cal B}_1$ or ${\cal B}_2$. 
Alice and Bob publicly discuss the bases chosen, discarding all the 
dits where they have selected different bases (sifting). 

Generalizing the ideas presented in \cite{BBM}, 
the aforementioned state-distribution process can be viewed 
as follows \cite{NA,C-d,BKBGC,AGS}.    
Alice prepares each of $N\gg 1$ entangled-qudit pairs in the 
maximally entangled state $\ket{\Psi_{00}}$. Thereby, the 
generalized maximally entangled states in the Hilbert space 
of two distinguishable qudits 
$\mathbb{C}_{\rm A}^d\otimes\mathbb{C}_{\rm B}^d$ are defined 
as $\ket{\Psi_{mn}} := \sum_{j\in\field{d}}\ket{j_{\rm A}}\otimes 
\Er{mn}^{\rm (B)}\ket{j_{\rm B}}/\sqrt{d}$,
where from now on the subscripts A and B refer to Alice and Bob, 
respectively \cite{C-d,BKBGC,AGS,ADGJ,MDN}.
Alice keeps half of each pair and submits the other 
half to Bob after having applied at random and independently,  
a unitary transformation chosen from the set $\{\unity, \Hr\}$. 
As soon as Bob receives the particles, he acknowledges the fact 
and applies at random $\unity$ or $\Hr^{-1}$ on each qudit independently. 
Alice reveals the sequence of operations she performed and   
all the pairs which involve different operations on the 
transmitted qudit are discarded. This is the associated 
entanglement-based (EB) version of the $2d$-state QKD protocol 
and offers many advantages, in particular with respect to security issues 
and error tolerance.  

The second stage of the QKD protocol is the verification test 
(also called signal-quality test) which we discussed in detail elsewhere \cite{NA}. 
In this stage, the two legitimate users sacrifice part of their (quantum) 
signal in order to quantify the eavesdropping rate during the transmission stage. 
More precisely, after a random 
permutation of their sifted (qu)dit pairs, Alice and Bob randomly select 
a sufficiently large number of them and determine their average 
error probability (disturbance). If as a result of a noisy quantum channel 
(from now on all the noise in the channel is attributed to eavesdropping)   
the estimated disturbance is too high, the protocol is aborted. 
Otherwise, Alice and Bob proceed to the key-distillation phase which 
will be discussed in detail in the following section.    

At any rate, it is always worth keeping in mind that the success of 
the verification test (and thus security) relies on two key points. 
First, an eavesdropper does not now in advance which qudit-pairs will be chosen 
for quality checks and which qudit-pairs will contribute to the final key. 
Second, any joint eavesdropping attack can be reduced to a 
classical (probabilistic) cheating strategy for which classical 
sampling theory can be safely applied 
\cite{GL,C-d,LC,SP}. 

In particular, the action of the quantum channel can be regarded as a 
Pauli one \cite{GL,C-d}. At the end of the distribution stage 
of the $2d$-state protocol, each transmitted qudit may have undergone any 
of the $d^2$ possible types of errors $\Er{mn}$. 
Let $p_{mn}$ denote the rate (probability) of errors of the 
form $\Er{mn}$ in the particles shared between 
Alice and Bob, with 
\bea
\sum_{m,n\in\field{d}}p_{mn}=1.
\label{norm}
\eea
In general, any symmetries underlying the QKD protocol under 
consideration may imply additional constraints on $p_{mn}$.  
For the protocols under consideration, both Fourier-dual bases are 
used at random and independently on each qudit-pair during the transmission. 
Moreover, the choices of the bases are not 
known to an eavesdropper, and they are publicly announced only after 
all the particles are in Bob's possession. Thus, as a result of the 
symmetry between the two bases, the quantum channel 
connecting Alice and Bob yields different sets of identical 
error-probabilities \cite{NA}. 
In particular, we have that 
\bea
p_{mn}=p_{n,d-m}=p_{d-m,d-n}=p_{d-n,m},\quad \forall\, m,n\in\field{d}.
\label{SymBases}
\eea 
Note that in highly symmetric protocols, the corresponding symmetry between 
all $(d+1)$ mutually unbiased bases leads to a depolarizing quantum channel 
with $p_{mn}=p_{01}$ for all $(m,n)\neq(0,0)$ \cite{C-d}.

In view of the symmetries (\ref{SymBases}), the estimated disturbance during 
the verification test is given by \cite{NA}
\bea
D=\sum_{m\in\field{d}^*}p_{m0}
+\sum_{m\in\field{d}^*}\sum_{n\in\field{d}^*}p_{mn},
\label{estD}
\eea
where $\field{d}^*:=\field{d}\backslash\{0\}$. This estimated error rate 
should not be confused with the so-called quantum-channel (overall) 
error rate $Q=1-p_{00}$, which is not estimable in a typical verification 
test of a P\&M $2d$-state QKD protocol. 

At this point, we have all the necessary formalism and we turn 
to investigate the error tolerance of $2d$-state P\&M protocols. 

\section{Analysis of the two-way key distillation}
\label{secII}
Throughout this work we focus on the GL2KD procedure in the context 
of which the highest tolerable error rates have been reported 
for various P\&M QKD schemes \cite{GL,C-2,C-d}. 
Our purpose is to investigate the conditions 
under which an insecure quantum channel allows the distillation 
of a secret key in the context of $2d$-state QKD protocols and 
the GL2KD procedure. Such an analysis can be performed conveniently 
in the EB version of the protocols we described in the previous 
section and adopt from now on. 
We will close this section with the reduction of the EB scheme to 
a P\&M one. 

\subsection{Dit-flip error rejection (DER)}
As any other key-distillation process, the GL2KD has two stages  
\cite{GL,C-2,C-d}. The first stage 
is a typical two-way entanglement purification with LOCC2 
\cite{ADGJ,MDN,DEJ,BDSW}. More precisely, in order to reduce 
the dit-flip-error rate in their signal Alice and Bob 
apply a number of D-steps. In each D-step, they form tetrads 
of particles by randomly pairing up their qudit-pairs.  
Then, within each tetrad of particles they apply a bilateral 
exclusive OR (BXOR) operation. Specifically, Alice and Bob individually 
apply to their halves the unitary operation 
\bea
{\rm XOR}_{{\rm c}\to{\rm t}}: \ket{x}_{\rm c}\otimes 
\ket{y}_{\rm t}\mapsto\ket{x}_{\rm c}\otimes 
\ket{x-y}_{\rm t},
\eea
where ${\rm c}$ and ${\rm t}$ denote the control and target qudit, respectively. 
Subsequently, they measure their target qudits in the computational basis  
and compare their outcomes. The control qudit-pair is kept if and only if 
their outcomes agree, while the target pair is always discarded.

In general, this procedure is repeated many times 
(many rounds of D-step) until the dit-flip-error rate 
in the surviving qudit-pairs is sufficiently low to 
guarantee an arbitrarily small total error rate 
at the end of the key-distillation protocol. We are 
going to make this statement more precise later on. 
For the time being, we turn to analyze the effect 
of the D-steps on the signal shared between Alice and Bob.     

Following \cite{GL,C-2,C-d}, our analysis will be based on classical 
probability arguments since any eavesdropping attack can be reduced to 
a classical probabilistic one. In particular, let 
$S=\{p_{mn}|~m,n\in\field{d}\}$ be the set of error rates 
(error-probability distribution) at the beginning of 
DER (i.e., at the end of the first stage of the QKD protocol). 
It has been shown \cite{MDN} that the effect of $k$ rounds of 
D-step (with $k\in\PosInt$) on 
$S$ can be identified by a mapping 
${\cal D}_k: S\mapsto S_k$,  
where $S_k=\{p_{mn}^{(k)}|~m,n\in\field{d}\}$ 
and   
\bea
p_{mn}^{(k)} &=& \frac{
\sum_{l\in\field{d}}\Phi(-n\cdot l)
\left [ \sum_{j\in\field{d}}\Phi(l\cdot j)~p_{mj}
\right ]^{2^k}}{d\sum_{i\in\field{d}}\left (
\sum_{j\in\field{d}}p_{ij}\right )^{2^k}}.
\label{map1}
\eea
One can readily check that by setting $d=2$, this mapping reduces to 
the well-known mapping for qubit-based protocols \cite{GL,C-2}.
 
Clearly, $p_{mn}^{(k)*}=p_{mn}^{(k)}$ since the summations in 
Eq. (\ref{map1}) run over all the finite field $\field{d}$. 
Furthermore, for the same reason, Eq. (\ref{map1}) can be rewritten as 
\bea
p_{mn}^{(k)} &=& 
\frac{
\left [ C(m)\right ]^{2^k}+
\sum_{l\in\field{d}^*}\Phi(-l\cdot n)
\left [
A(m,l)
\right ]^{2^k}}
{d\left [1 +
\sum_{l\in\field{d}^*}
\left [
C(l)
\right ]^{2^k}\right ]},
\label{map2}
\eea 
where 
\begin{subequations}
\label{AB-par}
\bea
A(m,l)&=&
\frac{\sum_{j\in\field{d}} \Phi(l\cdot j)p_{m j}}{\sum_{j\in\field{d}} p_{0j}},
\\
C(m)&=&\frac{\sum_{j\in\field{d}} p_{m j}}{\sum_{j\in\field{d}} p_{0j}},
\eea
\end{subequations}
for $m,l\in\field{d}$. 

From now on we restrict ourselves to estimated disturbances $D< D_{\rm th}$, since 
for $D \geq D_{\rm th}$ Alice and Bob do not share provable entanglement 
\cite{NA,GL,C-2}. Furthermore, for $D< D_{\rm th}$ we also have   
\bea
\sum_{n\in\field{d}}p_{0n}>
\sum_{n\in\field{d}}p_{mn}\quad\forall\, m\in\field{d}^*, 
\label{dist_id}
\eea
which implies that 
$0\leq C(m)< 1,\forall~m\in\field{d}$.  
Besides, a necessary condition for $0\leq p_{mn}^{(k)}\leq 1$ 
after many rounds of D-step is $|A(m,l)|<1$, for all 
$m,l\in\field{d}$. Thus, as $k\to\infty$, we have 
$|A|^{2^k}\to 0$ 
and $|C|^{2^k}\to 0$ which imply that $p_{0n}^{(k)} \to 1/d$ and 
$p_{mn}^{(k)} \to 0$,  for $m,n\in\field{d}$ and $m\neq 0$. 
In other words, the main effect of DER on the surviving particles 
shared between Alice and Bob is to reduce errors of 
the form $\Er{mn}$ with $m\neq 0$, while increasing the rate of pure phase 
errors of the form $\Er{0n}$ with $n\neq 0$. 

In particular, let 
\begin{subequations}
\label{tot_Rk}
\bea
R_{\rm D}^{(k)}=\sum_{m\in\field{d}^*}\sum_{n\in\field{d}}p_{mn}^{(k)}
\eea 
and 
\bea
R_{\rm P}^{(k)}=\sum_{m\in\field{d}}\sum_{n\in\field{d}^*}p_{mn}^{(k)}\equiv
\sum_{n\in\field{d}^*}q_{n}^{(k)}
\eea 
\end{subequations}
be the total dit-flip- and phase-error rates after $k$ rounds of D-step, respectively. 
As $k\to\infty$, $R_{\rm D}^{(k)}\to 0$ whereas $R_{\rm P}^{(k)}\to (d-1)/d$. 
We must therefore have a closer look at the corresponding individual phase-error 
rates $q_{n}^{(k)}$ which, using Eq. (\ref{map2}),  are given by 
\bea 
q_{n}^{(k)}=\sum_{m\in\field{d}}p_{mn}^{(k)}&=&
\frac{1}{d}+\frac{\xi_n^{(k)}}{d
\left [1+\chi^{(k)}\right ]}
\label{qD2}
\eea
for all $n\in\field{d}$, where 
\begin{subequations}
\label{small-par}
\bea
\xi_n^{(k)}&=&\sum_{m\in\field{d}}\sum_{l\in\field{d}^*}\Phi(-l\cdot n)\left[
A(m,l)\right ]^{2^k},\label{small-par-xi}\\
\chi^{(k)}&=&\sum_{m\in\field{d}^*}\left[C(m)\right ]^{2^k}.
\eea
\end{subequations} 
Clearly, the parameters $\xi_n^{(k)}$ and $\chi^{(k)}$ also take arbitrarily 
small values as $k\to\infty$, since $|A|^{2^k}\to 0$ and $|C|^{2^k}\to 0$. 

{\em Observation 1}. The phase-error rates after $k$ rounds of D-step satisfy the 
inequality 
\bea
q_{0}^{(k)}>q_{n}^{(k)} \quad\forall n\in\field{d}^*,
\label{mvec_nc}
\eea
where $q_0^{(k)}$ is the no-phase-error probability.

{\em Proof}. First of all, recall that throughout this work we assume 
prime dimensions only. Starting from Eq. (\ref{qD2}), we have to show that 
$\xi_0^{(k)} > \xi_n^{(k)}$, for all $n\neq 0$. 
Using the symmetry condition (\ref{SymBases}), 
Eq. (\ref{small-par-xi}) reads
\bea
\xi_n^{(k)}&=&
\frac{2\sum_{m=0}^{\lfloor d/2 \rfloor}\sum_{l=1}^{\lfloor d/2 \rfloor}
\cos(l\cdot n)T(m,l)}{\left [\sum_{j\in\field{d}}p_{0j}\right ]^{2^k}}\quad
\forall n\in\field{d},\nonumber\\
\label{xi_n_Eq2}
\eea  
where all $T(m,l)$ are real and positive. In particular, we have that  
\bea
T(0,l)&=&\left [p_{00}+2\sum_{j=1}^{\lfloor d/2\rfloor}\cos(l\cdot j)p_{0j}
\right ]^{2^k},
\nonumber \\
T(m,l)&=&2\Re\left \{
\left [\sum_{j\in\field{d}}\Phi(l\cdot j)p_{mj}\right ]^{2^k}\right \},
\quad {\rm for}\, m\neq 0.
\nonumber
\eea
where $\Re(x)$ denotes the real part of $x$.
In view of  Eq. (\ref{xi_n_Eq2}),  Eq. (\ref{mvec_nc}) now follows immediately 
from the inequality $\xi_0^{(k)} > \xi_n^{(k)}$ as a consequence of the fact that  
$\cos(x)<1,\, \forall\, x\in\field{d}^*$. A similar but more involved calculation 
can be performed if $d$ is a prime power. \hfill $\blacksquare$

\subsection{Phase error correction (PEC)} 
Assume now that Alice and Bob have applied a DER process involving many 
$(k\gg 1)$ rounds of D-step. As we have just discussed, at 
this point the dit-flip-error rate in their surviving pairs will 
be negligible (i.e., $p_{mn}^{(k)}\simeq 0$ for $m\neq 0$), 
whereas the 
phase-error rate has possibly increased.  
It is therefore reasonable that the second stage of the GL2KD 
(usually called privacy amplification) deals with phase error 
correction (PEC) \cite{GL,C-2,C-d}. 

In general, at the beginning of the PEC we have a $d$-ary asymmetric channel 
with respect to phase errors. In particular, we have $(d-1)$ possible 
phase errors with corresponding probabilities (rates) $q_{n}^{(k)}$ 
given by Eq. (\ref{qD2}). To correct the phase errors, Alice and Bob apply 
an $[r,1,r]_d$ repetition code with a relative majority-vote decoding 
\cite{ECC-book}.
The key point is that, according to 
inequality (\ref{mvec_nc}), the necessary condition \cite{ECC-book} 
for such an error correction to work is satisfied at the end of the DER process. 

For the sake of completeness, let us briefly summarize the main steps of the PEC 
procedure \cite{GL,C-2,C-d}. Alice and Bob randomly divide their qudit-pairs into 
sets (blocks), each containing $r$ qudit-pairs. Within each block, they perform a 
discrete Fourier transform $\Hr_{\rm A}\otimes\Hr_{\rm B}$ on each pair. 
Subsequently, they perform a sequence of $(r-1)$ BXOR operations with the 
same control pair (say the first one) and targets each one of the remaining pairs. 
For each target pair, they measure their corresponding halves and estimate the 
parity of their outcomes. Finally, they apply $\Hr^{-1}_{\rm A}\otimes\Hr^{-1}_{\rm B}$ 
on the control pair and Bob performs $\Er{0s}$ on his control-qudit, where 
$s\in\field{d}$ is the parity corresponding to the relative majority of their $(r-1)$ 
outcomes. If the relative majority of the outcomes is ambiguous, Bob applies $\Er{00}$.  
In this way, each block may result in one phase-error-free qudit-pair at most. 
 
Our task now is to investigate the effect of such a PEC process on the 
signal shared between Alice and Bob.  Let us denote by $p_{mn}^{\rm P}$ the various 
error rates in the remaining qudit-pairs at the end of the process. 
We are mainly interested in the corresponding total dit-flip- and phase-error rates.

\subsubsection{Phase-error rate} 
Let us start with the estimation of an upper bound on the total 
phase-error rate $R_{\rm P}\equiv\sum_{m}\sum_{n\neq 0} p_{mn}^{\rm P}$ 
of the signal at the end of PEC. 
We are basically interested in the limit of large block-lengths $r$, 
that is in the limit of a large number of distributed qudit-pairs. 

Before we proceed further, it is worth noting that the problem under 
consideration belongs to a well known class of stochastic processes,  
the so-called occupancy problems or Balls-and-Bins experiments. 
In this picture, our problem can be viewed as a probabilistic experiment 
where $r$ balls (qudit-pairs) are randomly distributed among 
$d$ different (error-)bins. This class of problems is fundamental to the 
analysis of 
randomized algorithms and has been extensively studied in the literature 
(e.g., see \cite{brics,SSS,C-H-book-1}). A particularly useful result in this context 
are the so-called Chernoff-Hoeffding bounds \cite{C-H-cite} which are 
basically large-deviation estimates. In general, these bounds are applicable 
to sums of negatively associated, identically distributed random variables. 
Their precise derivation can be found in various papers and 
standard textbooks (e.g., see \cite{SSS,C-H-cite,C-H-book-1,C-H-book-2}). 

{\em Observation 2}. The phase-error rate in the surviving pairs 
at the end of PEC satisfies the condition 
\bea
R_{\rm P}\leq 
\sum_{n\in\field{d}^*}\left[1-\left (
\sqrt{q_0^{(k)}}-\sqrt{q_n^{(k)}}~
\right )^2\right ]^{r}.
\label{RPbound}
\eea

{\em Proof.} 
Clearly, we have that $R_{\rm P}$ is upper bounded by the probability of 
failure for the repetition code $P_{\rm fail}$. It suffices therefore, to estimate 
an upper bound on $P_{\rm fail}$. 

As we mentioned before, PEC is applied on a particular asymmetric channel 
with phase-error rates $q_0 > q_j$ for all $j\neq 0$ 
(to simplify notation throughout this proof we write $q_j$ instead of 
$q_j^{(k)}$). 
Let us denote by $\eta_j$ the total number of qudit-pairs within a block of 
length $r$ suffering from phase errors of the form $\Er{mj}$, with $m\in\field{d}$. 
Clearly, majority voting fails only if $\eta_j>\eta_0$ for some $j\neq 0$, 
where $\eta_0$ denotes the number of error-free pairs in the block. 
For asymmetric channels satisfying Eq. (\ref{mvec_nc}), this may occur for sufficiently 
large deviations of $\eta_j$ from their mean values. 
In particular, we expect for the failure probability of the majority-vote decoding, 
\bea
P_{\rm fail}\leq P\left [ 
\bigvee_{j\in\field{d}^*} \left (\eta_j\geq \eta_0\right )\right ]
\leq \sum_{j\in\field{d}^*} P\left (\eta_j\geq \eta_0\right ).\nonumber\\
\label{Bonf}
\eea
where $\bigvee$ is the logical OR operator. 
The next step now is to upper bound each of the probabilities 
$P\left (\eta_j\geq \eta_0\right )$ appearing in the last summation.  

Let us focus on a particular term, say $P\left (\eta_i\geq \eta_0\right )$. 
We will work with the radom variables 
$\eta_i$, $\eta_0$ and $\eta_{\rm rest}$, where $\eta_i+\eta_0+\eta_{\rm rest}=r$ and 
$\eta_{\rm rest}=\sum_{j\not{\in}\{0,i\}} \eta_j$. 
Accordingly, the corresponding probability distribution of interest is 
$(q_0, q_i, q_{\rm rest})$ with $q_i+q_0+q_{\rm rest}=1$. Obviously,  
$(\eta_0, \eta_i, \eta_{\rm rest})$ have a trinomial distribution which is given by  
\bea
P(\eta_0,\eta_i,\eta_{\rm rest})=\sum_{\eta_{\rm rest}=0}^r\binom{r}{\eta_{\rm rest}} q_{\rm rest}^{\eta_{\rm rest}} 
\left [ 
\sum_{\eta_i=0}^{r_i}\binom{r_i}{\eta_i}q_0^{\eta_0}q_{i}^{\eta_i} \right ],
\nonumber
\eea
where $r_i=\eta_0+\eta_i=r-\eta_{\rm rest}$. Introducing the new normalized 
probabilities $\tilde{q}_l=q_l/(q_0+q_i)$ with $l\in\{0,i\}$, 
the trinomial distribution can be rewritten as 
\bea
P(\eta_0,\eta_i,\eta_{\rm rest})&=&\sum_{\eta_{\rm rest}=0}^r\binom{r}{\eta_{\rm rest}}
q_{\rm rest}^{\eta_{\rm rest}} (q_0+q_i)^{r-\eta_{\rm rest}}\nonumber\\
&&\times\left [ 
\sum_{\eta_i=0}^{r_i}\binom{r_i}{\eta_i}\tilde{q}_0^{\eta_0}
\tilde{q}_{i}^{\eta_i} \right ].
\nonumber
\eea 
Note now that the expression in the brackets is the well 
known binomial distribution involving the two events of interest, 
i.e., the event of phase-error $i$, and the event of no-phase-error. 
In particular, for a given $\eta_{\rm rest}$ the probability that $\eta_i\geq\eta_0$ 
is given by 
\bea
P\left (\eta_i\geq \eta_0~|~\eta_{\rm rest} \right )&=&
\sum_{\eta_i=\lceil r_i/2 \rceil}^{r_i}
\binom{r_i}{\eta_i}
\tilde{q}_0^{\eta_0}\tilde{q}_{i}^{\eta_i}\nonumber\\
&\leq&
\left (4\tilde{q}_0\tilde{q}_{i}\right )^{r_i/2}=
\left [\frac{4 q_0 q_i}{(q_0+q_i)^2} \right ]^{r_i/2}.
\nonumber
\eea
The above inequality is the well-known Chernoff-Hoeffding bound 
for the binomial distribution \cite{C-H-book-2}, 
which also applies here since $q_0>q_i$ $\forall\, i\in\field{d}^*$. 
Thus, in total we have   
\begin{widetext}
\bea
P\left (\eta_i\geq \eta_0\right )&=&
\sum_{\eta_{\rm rest}=0}^r\binom{r}{\eta_{\rm rest}} 
q_{\rm rest}^{\eta_{\rm rest}} (1-q_{\rm rest})^{r-\eta_{\rm rest}}
P\left (\eta_i\geq \eta_0~|~\eta_{\rm rest} \right )
\nonumber\\
&\leq& 
\sum_{\eta_{\rm rest}=0}^r\binom{r}{\eta_{\rm rest}} 
q_{\rm rest}^{\eta_{\rm rest}} (1-q_{\rm rest})^{r-\eta_{\rm rest}}
\left [\frac{4 q_0 q_i}{(q_0+q_i)^2} \right ]^{(r-\eta_{\rm rest})/2}.
\label{p-ineq}
\eea
\end{widetext}
Finally, given that $R_{\rm P}\leq P_{\rm fail}$, 
inequality (\ref{RPbound}) is obtained 
from the condition (\ref{Bonf}), by using inequality 
(\ref{p-ineq}) and the identity 
$\sum_{a=0}^r\binom{r}{a}p^a(1-p)^{r-a}x^{r-a}=
\left[ p+(1-p)x\right ]^r$. \hfill $\blacksquare$

According to observation 2, 
the phase-error rate in the signal after PEC decreases exponentially in the 
block-length $r$. If we are not interested on a tight upper bound on $R_{\rm P}$, 
we may upper-bound the right-hand side of this condition as follows 
\bea
R_{\rm P}&\leq& \sum_{n\in\field{d}^*}\left[1-\left (
\sqrt{q_0^{(k)}}-\sqrt{q_n^{(k)}}~
\right )^2\right ]^{r}\nonumber \\
&\leq&(d-1)\left[1-\left (
\sqrt{q_0^{(k)}}-\sqrt{q_{\tilde{n}}^{(k)}}~
\right )^2\right ]^{r}.
\label{RPbound2}
\eea
where $q_{\tilde{n}}^{(k)}=\max\left \{q_{n}^{(k)}~\left 
|~\right.n\in\field{d}^*\right\}$, while equality in the latter part holds 
if and only if 
$q_{n}^{(k)}=q_{\tilde{n}}^{(k)}$, $\forall\,n\in\field{d}^*$. 
Although this last step is not at all 
necessary, it considerably simplifies the subsequent notation and discussion.

Recall now that the quantities $\xi_{n}^{(k)}$ and $\chi^{(k)}$ 
become arbitrarily small as $k\to\infty$. Thus, in view of Eq. (\ref{qD2}), 
Eq. (\ref{RPbound2}) may further simplified to  
\bea
R_{\rm P}&\leq& (d-1)\left[1- ~\frac{\left ( 
\xi_0^{(k)}-
\xi_{\tilde{n}}^{(k)}
\right )^2}{4d}+O\left (3 \right )\right ]^{r},
\nonumber
\eea
where $O(3)$ denotes third-order terms in $\xi_{\tilde{n}}^{(k)},\,\chi^{(k)}$ 
and $\xi_{0}^{(k)}$. 
Inclusion of such higher-order terms may only lead to negligible corrections 
in the argument of the exponent. At any rate, the phase-error rate will always 
be upper-bounded by a quantity which decreases exponentially fast in $r$.  
Alternatively, using the inequality $(1 - x)^r\leq \exp(-rx)$ for all $x<1$, 
we obtain 
\bea
R_{\rm P}&\leq& (d-1) \exp
\left [-r \frac{\left ( \xi_0^{(k)}-
\xi_{\tilde{n}}^{(k)} \right)^2}{4d}\right ].
\label{RP-bound3}
\eea
We turn now to estimate the corresponding dit-flip-error rate 
in the signal.
 
\subsubsection{Dit-flip-error rate}
As we mentioned before, the PEC involves $(r-1)$ BXOR gates 
in the complementary basis. During these gates the dit-flip errors propagate 
backwards from the target to the control qudit. 
As a result, at the end of the PEC the dit-flip-error 
rate in the remaining particles increases by at most $r$ times 
(the control qudit-pair itself may initially suffer from a dit-flip-error), i.e.,   
\bea
R_{\rm D}\equiv \sum_{m\in\field{d}^*}\sum_{n\in\field{d}}p_{mn}^P
&\leq& r\sum_{m\in\field{d}^*}\sum_{n\in\field{d}}p_{mn}^{(k)}.
\label{RD-bound}
\eea

According to the preceding discussion the net effect of the PEC is 
to reduce any phase errors of the form $\Er{mn}$ with $n\neq 0$, 
while possibly increasing dit-flip errors of the form $\Er{m0}$ 
with $m\neq 0$. Thus, at first site, the whole situation seems to be a 
vicious circle since PEC tends to destroy what was achieved in DER 
and vice-versa. A way out of this stumbling block relies on the judicious 
combination of DER and PEC. 

\subsection{A judicious combination of DER and PEC}
For a given $2d$-state protocol (i.e., for a fixed $d$) 
Alice and Bob agree in advance upon a fixed and arbitrarily small security parameter  
$\epsilon>0$. They apply many rounds ($k\gg 1$) of D-step, until there exists an integer 
$r>0$ such that a single application of the PEC will bring the quantum-channel error 
rate in the finally surviving pairs to values below $\epsilon$. 
Clearly, the protocol has to be aborted if the estimated integer $r$ exceeds   
the number of remaining pairs immediately after the DER procedure. 
More precisely, at the 
end of DER, Alice and Bob may choose the block length for the repetition code to be 
\bea
r\approx
\frac{\epsilon}{2\sum_{m\in\field{d}^*}\sum_{n\in\field{d}}p_{mn}^{(k)}}=
\frac{\epsilon}{2}~\bigg(1+\frac{1}{\chi^{(k)}}\bigg)\geq 
\frac{\epsilon}{2\chi^{(k)}}.
\label{r_opt}
\eea  
Note that for this particular choice of the block-length, $r\to\infty$ as $k\to\infty$. 
   
The key point now is that for such a choice of $r$, the overall channel error rate 
$Q=1-p_{00}^{\rm P}$ can be upper-bounded as follows  
\bea
Q&\leq&R_{\rm D}+R_{\rm P}\nonumber\\
&\leq&\frac{\epsilon}{2}+(d-1)\exp\left [-\frac{\epsilon}{8}
\frac{\left (\xi_0^{(k)}-\xi_{\tilde{n}}^{(k)} \right )^2}{d\chi^{(k)}} \right],
\eea
where inequalities  (\ref{RD-bound}) and  (\ref{RP-bound3}) have been used.
Thus, for any given dimension of the information carriers, 
$Q<\epsilon$ provided that 
\bea
\frac{\left [\xi_0^{(k)}-\xi_{\tilde{n}}^{(k)}\right ]^2}{d\chi^{(k)}}>
\frac{8}{\epsilon}\ln\left [\frac{2(d-1)}{\epsilon}\right ],
\label{distil-cond-as}
\eea 
As long as $Q<\epsilon$, Alice and Bob share a number of 
nearly perfect pairs whose fidelity with respect to the ideal state 
$\ket{\Psi_{00}}$ is exponentially close to one. The final key can  
then be obtained by measuring each pair separately along the standard 
basis, and the information that an eavesdropper may have on it, 
is also upper bounded by the security parameter $\epsilon$. 

The condition  (\ref{distil-cond-as}) is a sufficient condition for secret-key 
distillation in the context of $2d$-state QKD protocols using two Fourier-dual 
bases. In particular, it determines the error rates  
which can be tolerated by such protocols using a GL2KD procedure. 
From that point of view, it is a generalization of the corresponding 
condition for fully symmetric qudit-based protocols obtained by Chau \cite{C-d}.

Unfortunately, the number of independent parameters in inequality (\ref{distil-cond-as}) 
scales quadratically with $d$, and thus an analytical (or even numerical) solution 
becomes rather difficult for $d>3$. Hence, in order to obtain an analytic expression for the 
tolerable error rate for arbitrary dimensions we had to resort to isotropic 
quantum channels. The related results will be discussed in detail in 
Sec.  \ref{Sec-Iso}. For the time being we close this section by summarizing the 
main points in the reduction of the EB version of the $2d$-state QKD protocol to 
a P\&M one.  
  
\subsection{Reduction to a P\&M QKD scheme}
In general, not every EB QKD protocol can be reduced to a P\&M one. 
The main difficulty appears in the reduction of the underlying 
quantum key-distillation procedure to a purely classical one. 
The advantage of the GL2KD is that by construction it allows for such 
a reduction \cite{GL}. 

The reduction of the EB $2d$-state QKD protocol to a P\&M one, 
which tolerates precisely the same error rates, follows the 
same steps as for other protocols \cite{GL,C-d,SP}.  
Here, for the sake of completeness, we would like to summarize the four 
cornerstones of such a reduction.
First, during the distribution stage, Alice can measure all the halves 
of the pairs before sending the other halves to Bob. This is equivalent to 
choosing a random dit-string and encoding each dit in the corresponding 
qudit-state, in one of the two Fourier-dual bases.  
Second, the XOR operation used in the quantum key-distillation 
procedure can be easily replaced by its classical analogue. 
Thus, the DER stage is immediately reduced to a classical error-rejection 
(advantage distillation) process. 
Third, the quantum circuit of the PEC can also 
be reduced to a classical one. Such a reduction relies on the fact 
that the sequence of gates applied independently by Alice and Bob in each block of $r$ 
qudits during PEC, i.e., 
$\Hr_{1}^{-1}\left (
{\rm XOR}_{1\to r}\ldots {\rm XOR}_{1\to 2}\right )\bigotimes_{j=1}^{r}\Hr_{j}$,
is equivalent to $\bigotimes_{j=2}^r\Hr_{j}^{-1} 
\left ({\rm XOR}_{r\to 1}^{(+)}\ldots {\rm XOR}_{2\to 1}^{(+)}\right )$. 
This equivalence follows by induction from the fact that 
for any two qudits, 
$\left (\Hr_{\rm c}^{-1}\otimes \unity_{\rm t} \right )
{\rm XOR}_{{\rm c}\to {\rm t}} \left (\Hr_{\rm c}\otimes \Hr_{\rm t} \right )
= \left ( \unity_{\rm c} \otimes \Hr_{\rm t}^{-1}\right )
{\rm XOR}_{{\rm t}\to {\rm c}}^{(+)}$, where 
${\rm XOR}_{{\rm c}\to{\rm t}}^{(+)}: \ket{x}_{\rm c}\otimes 
\ket{y}_{\rm t}\mapsto\ket{x}_{\rm c}\otimes 
\ket{x+y}_{\rm t}$. 
Finally, the last essential point in the reduction is the observation that the 
key-distillation procedure does not rely on phase information.   

The above steps lead to a P\&M $2d$-state QKD protocol with the  
distribution and the verification-test stages discussed in Sec. \ref{intro-2}.
The corresponding classical key-distillation stage of the protocol proceeds 
as follows \cite{GL,C-2,C-d}. 

{\bf DER:} Alice and Bob perform many rounds of D-step. In each round they randomly 
form tetrads of their dits. For each tetrad $j$, Alice announces the 
parity of her dits, i.e., she announces $X_{1}^{(j)}-X_{2}^{(j)}$, where 
$X_{i}^{(j)}$ denotes the $i$-th pair of tetrad $j$.  
Similarly, Bob announces the parity of his corresponding dits $Y_{1}^{(j)}-Y_{2}^{(j)}$.
One of the dit-pairs (say $X_{1}^{(j)}$ and $Y_{1}^{(j)}$) survives if and only if 
the announced parities agree. This process is 
repeated (many rounds of D-step), until there is an integer $r>0$ such that 
a single application of the following phase-error correction will bring the overall 
error rate in the remaining signal below $\epsilon$. The protocol is aborted if the estimated 
parameter $r$ exceeds the number of remaining dits. 

{\bf PEC:} In the classical PEC (which is essentially privacy amplification), 
Alice and Bob randomly divide their remaining dit-pairs into blocks each containing 
$r$ dit-pairs. Let us denote by $(X_{i}^{(j)},Y_{i}^{(j)})$ the $i$-th dit-pair in 
block $j$. Alice and Bob, replace each block by the parity of its dits, i.e., 
by $\sum_{i=1}^rX_{i}^{(j)}$ and 
$\sum_{i=1}^rY_{i}^{(j)}$, respectively.   
In this way, the final secret key essentially consists of the estimated 
parities for each one of the blocks. 

In closing, it has to be noted here that for a more efficient secret-key distillation 
the two legitimate users may follow the adaptive key-distillation procedure introduced 
by Chau \cite{C-2,C-d}. The main difference is that Alice and Bob do not apply many 
rounds of D-step and PEC in order to bring the overall error rate below the security 
parameter $\epsilon$. Instead, they simply adjust their DER and PEC in order to 
bring the overall error rate below, let us say $5\%$. From that point on, they  
switch to more efficient error-correction and privacy amplification using 
concatenated Calderbank-Shore-Steane codes.

\section{Isotropic quantum channels}
\label{Sec-Iso}
An isotropic channel is characterized by $p_{0j}=p_{j0}=p_{10}$ and 
$p_{ij}=p_{ji}=p_{11}$ for $i,j\in\field{d}^*$. 
It turns out that isotropy is an inherent property of the two-basis protocols 
using qubits (standard BB84) or qutrits \cite{NA}. However, in general for $2d$-state 
protocols using higher dimensions $(d>3)$, isotropy cannot be 
justified so easily, unless the quantum channel itself is 
isotropic (e.g., open-space quantum cryptography). 

The robustness and security of various QKD protocols under the assumption of 
isotropic eavesdropping has been extensively studied in the QKD literature 
\cite{BKBGC,AGS,PABM,PT,DKCK,CG-FGNP}. In particular, we know that  at any rate the isotropy 
assumption does not affect the threshold disturbance for secret-key distillation 
which, for $2d$-state protocols, is given by Eq. (\ref{ThdistilConst}) \cite{NA}. 
In this section, our purpose is to further analyze the sufficient condition 
for key distillation (\ref{distil-cond-as})  in the framework of isotropic 
quantum channels and derive an analytic expression for the tolerable error 
rate of $2d$-state QKD protocols.

Instead of isotropic channels, we may consider a slightly more general class of 
channels for which $p_{0j}\neq p_{j0}$, that is  
\bea
p_{mn}=\left (
\begin{array}{cccc}
p_{00} & p_{01} & \ldots & p_{01}\\
p_{10} & p_{11} & \ldots & p_{11}\\
\vdots & \vdots & \ddots & \vdots\\
p_{10} & p_{11} & \ldots & p_{11} 
\end{array}
\right ).
\label{mat_iso2}
\eea 
Given the normalization condition (\ref{norm}), such a channel involves three independent 
parameters and thus the derivation of an analytic expression for the tolerable error rate 
is possible. Moreover, by setting $p_{01}=p_{10}$ we can easily obtain the corresponding 
expressions for isotropic channels. 

\subsection{Tolerable error rates}
For channels satisfying Eq. (\ref{mat_iso2}), Eq. (\ref{map2}) yields for the probabilities 
after $k$ rounds of D-step
\bea
p_{00}^{(k)} &=& \frac{[p_{00}+(d-1)p_{01}]^{2^k}+(d-1)(p_{00}-p_{01})^{2^k}}{d~\Pi},\nonumber\\ 
p_{0n}^{(k)} &=& \frac{[p_{00}+(d-1)p_{01}]^{2^k}-(p_{00}-p_{01})^{2^k}}{d~\Pi},\nonumber\\ 
p_{m0}^{(k)} &=& \frac{[p_{10}+(d-1)p_{11}]^{2^k}+(d-1)(p_{10}-p_{11})^{2^k}}{d~\Pi},\nonumber\\ 
p_{mn}^{(k)} &=& \frac{[p_{10}+(d-1)p_{11}]^{2^k}-(p_{10}-p_{11})^{2^k}}{d~\Pi},\nonumber 
\eea 
where $\Pi=[p_{00}+(d-1)p_{01}]^{2^k}+(d-1)[p_{10}+(d-1)p_{11}]^{2^k}$.
In view of these relations,  the form (\ref{mat_iso2}) is invariant under D-steps since we have 
that $p_{0n}^{(k)}=p_{01}^{(k)}$, $p_{m0}^{(k)}=p_{10}^{(k)}$ and 
$p_{mn}^{(k)}=p_{11}^{(k)}$, $\forall\, m,n\neq 0$. Therefore, all the phase-error rates 
$q_{n}^{(k)}$ with $n\neq 0$, are equal 
at the end of DER and the corresponding quantum channel is therefore symmetric with respect 
to phase errors. 

As in the previous section, we may also introduce the parameters $A(m,n)$ and $C(m)$. 
In fact, for the particular class of channels under consideration $A(m,n)=A(m)$ for all 
$m\in\field{d}$ and 
\begin{subequations}
\label{AB_iso}
\bea
A(0)&=&\frac{p_{00}-p_{01}}{p_{00}+(d-1)p_{01}},\quad \\
A(m)&=&A(1)=\frac{p_{10}-p_{11}}{p_{00}+(d-1)p_{01}}\quad{\rm for}\,\, m\neq 0,\quad\\
C(m)&=&C(1)=\frac{p_{10}+(d-1)p_{11}}{p_{00}+(d-1)p_{01}}\quad{\rm for}\,\, m\neq 0,\quad
\eea 
\end{subequations}
while $C(0)=1$. 
To proceed further, we note that $A(m)=B(m)C(m)$, where 
\bea
B(m)=\frac{p_{m0}-p_{m1}}{p_{m0}+(d-1)p_{m1}}=B(1),
\label{Beq_iso}
\eea 
and $[B(m)]^{2^k}\to 0$, as $k\to\infty$.  
Thus, using Eqs. (\ref{AB_iso}) and  (\ref{Beq_iso}),
Eqs. (\ref{small-par}) can be simplified to
\begin{subequations}
\bea  
\xi_0^{(k)}&=&(d-1)\sum_{m\in\field{d}}
\left [A(m)\right ]^{2^k}, \\ 
\xi_{n}^{(k)}&=&-\sum_{m\in\field{d}}
\left [A(m)\right ]^{2^k}\quad{\rm for}\,\, n\neq 0, \\ 
\chi^{(k)}&=& (d-1)\left [C(1)\right ]^{2^k},
\eea
\end{subequations}
where  
\bea
\sum_{m\in\field{d}} [A(m)]^{2^k}&=&[A(0)]^{2^k}+
\sum_{m\in\field{d}^*} [B(m)]^{2^k}[C(m)]^{2^k}\nonumber\\
&=& [A(0)]^{2^k}+(d-1)[B(1)C(1)]^{2^k}.
\eea 
Accordingly, condition (\ref{distil-cond-as}) now reads
\bea
\frac{d\left \{   [A(0)]^{2^k}+(d-1)[B(1)C(1)]^{2^k}
\right \}^2}{(d-1)[C(1)]^{2^k}}>
\frac{8}{\epsilon}\ln\left [\frac{2(d-1)}{\epsilon}\right ],
\nonumber
\label{distil-cond-iso2}
\eea
or equivalently [setting $A=A(0)$, $B=B(1)$ and $C=C(1)$]
\bea
\frac{dA^{2^{k+1}}}{(d-1)C^{2^k}}+d(d-1)C^2 B^{2^{k+1}}
+2dA^{2^k}B^{2^k}> f(d,\epsilon),\nonumber\\
\label{distil-cond-iso2a}
\eea
where $f(d,\epsilon)=8~\epsilon^{-1}\ln\left [2(d-1)/\epsilon\right ]$.

Recall now that the positive quantities $A^{2^k}\to 0$, $C^{2^k}\to 0$ 
and $B^{2^k}\to 0$ for $k\to\infty$. 
Thus, inequality (\ref{distil-cond-iso2a}) can always be satisfied for any 
$k$ such that   
\bea
\frac{dA^{2^{k+1}}}{(d-1)C^{2^k}}> f(d,\epsilon).
\label{distil-cond-iso2b}
\eea
For a given dimension, this latter inequality defines the critical 
number of D-steps $k_{\rm c}$, such that for $k>k_{\rm c}$ inequality 
(\ref{distil-cond-iso2a}) is satisfied. In particular, solving 
(\ref{distil-cond-iso2b}) with respect to $k$ we obtain  
\bea
k_{\rm c} = \log_{2}\left \{
\frac{\ln \left [ (d-1) f(d,\epsilon)/d\right ]}{\ln(A^{2}/C)}\right \}.
\eea
This is a well defined quantity provided that $A^2>C$, i.e., for    
\bea
(p_{00}-p_{01})^2>[p_{10}+(d-1)p_{11}][p_{00}+(d-1)p_{01}].
\label{final_ineq}
\eea 
where Eqs. (\ref{AB_iso}) have been used.
The same inequality holds for isotropic channels but $p_{01}=p_{10}$. 
This is therefore a sufficient condition for secret-key distillation 
in the context of any $2d$-state QKD protocol under the assumption of 
isotropic quantum channels. In particular, it determines the error rates  
which can be tolerated by such protocols using a GL2KD process. 

Recall now that according to Eq. (\ref{estD}) the estimated disturbance for 
the isotropic channel is $D=[1-p_{00}+(d-1)^2p_{11}]/2$.
Moreover, due to the normalization condition (\ref{norm}), 
inequality (\ref{final_ineq}) actually involves two independent 
parameters (say $p_{00},\,p_{11}$).  Thus, estimating the values of 
$p_{00}$ which satisfy it, we obtain the tolerable error 
rate (disturbance) which depends on both $d$ and $p_{11}$, 
i.e., $D_{\rm 2CC}(d, p_{11})$. 
In fact, we find that $D_{\rm 2CC}(d, p_{11})$ increases monotonically with 
respect to $p_{11}$. Hence, the worst-case scenario 
(from Alices's and Bob's point of view) corresponds to $p_{11}=0$ for which we 
obtain for the tolerable disturbance 
\bea
D_{\rm 2CC}(d)&=&\frac{1-p_{00}}2=\frac{2(d-1)}{4d-1+\sqrt{1+4d}},
\eea 
where $D_{\rm 2CC}(d)=D_{\rm 2CC}(d, p_{11}=0)$. Given a particular dimension of the 
information carriers (i.e., a specific $2d$-state protocol), the GL2KD procedure 
enables Alice and Bob to generate a provably secure key whenever the estimated 
disturbance is below $D_{\rm 2CC}(d)$ or else, the quantum channel 
error rate is below $2D_{\rm 2CC}(d)$.  
\\
\\  
\begin{figure}[h]
\resizebox{0.75\columnwidth}{!}{%
  \includegraphics{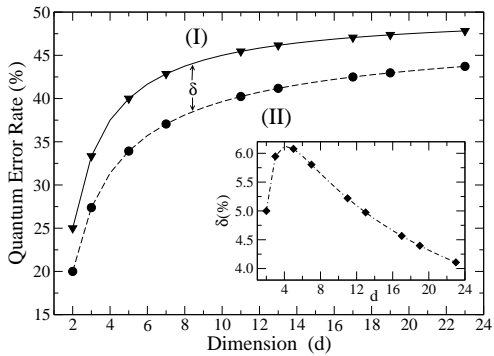}
}
\caption{$2d$-state QKD protocols : The tolerable error rate 
$D_{\rm 2CC}$ (dashed line) and its theoretical upper bound 
$D_{\rm th}$ (solid line) as functions of the dimension $d$.  
Secret-key distillation is impossible in the regime (I), while 
it may be possible for error rates below $D_{\rm th}$. 
In the regime (II) a secret key can be distilled by means of the 
key-distillation procedure considered here. 
Inset: The gap between the two regimes $\delta(d)=D_{\rm 2CC}-D_{\rm th}$  
is plotted  as a function of the dimension. The symbols 
(triangles, circles and squares) correspond to prime dimensions.
}
\label{Dth:fig}
\end{figure}

\subsection{Discussion}
The tolerable disturbance $D_{\rm 2CC}$ and its theoretical upper bound $D_{\rm th}$ 
are plotted as functions of the dimension $d$, in Fig. \ref{Dth:fig}.
First of all, we see that $D_{\rm 2CC}(d)<D_{\rm th}$ for all $d$. 
Actually, the difference between the two bounds $\delta(d)\equiv D_{\rm th}-D_{\rm 2CC}$ 
scales with dimension as 
\bea
\delta(d)=\frac{(d-1)\left (-2+\sqrt{1+4d}\right )}{2d(4d-3)},
\eea
and is also plotted in the inset of Fig. \ref{Dth:fig}.  
It is also worth noting that $\delta$ increases as we go from qubits $(d=2)$  
to qutrits $(d=3)$. It reaches its maximum value around $d=4$ (i.e., for quatrits) and  
decreases monotonically for higher dimensions. Note that the same behavior also appears 
in the case of $(d+1)$-basis protocols \cite{C-d}. Moreover, as $d\to\infty$, we have 
that 
\[
D_{\rm 2CC}(d)\approx \frac{1}{2} - \frac{1}{4\sqrt{d}},
\]
while $\delta(d)\approx 1/4\sqrt{d}$. In other words, we see that the the tolerable 
error rate for the $2d$-state QKD protocols approaches its theoretical upper 
bound as $1/\sqrt{d}$ for $d\to\infty$. This is in contrast to the $(d+1)$-basis 
protocols where the corresponding asymptotic behavior scales with dimension as $1/d$. 

A special case of the isotropic channel we have just considered is the so-called depolarizing 
channel for which $p_{mn}=p_{01}$ for $(m,n)\neq (0,0)$. In this case, 
condition (\ref{final_ineq}) reduces to 
Eq. (36) of Ref. \cite{C-d} i.e., 
\[
(p_{00}-p_{10})^2>d~p_{10}\left [p_{00}+(d-1)p_{10}\right ].
\]
Note also that for $d=2$ we recover the well-known tolerable 
error rate of the standard BB84 protocol, i.e., $D_{\rm 2CC}(2)=20\%$ \cite{C-2,RA}.

In closing, it is worth noting that condition (\ref{final_ineq}) can also be 
obtained by generalizing the ideas of Ref. \cite{RA} to higher dimensions.  
More precisely, let us define the characteristic exponent 
$r_{\rm ch}^{(d)}\in\Real$ with the defining property that there exists an 
$\alpha>0$ such that  
\bea
\lim_{k\to\infty}\frac{R_{\rm D}^{(k)}}{\left (
\frac{d-1}d-R_{\rm P}^{(k)}\right )^{r_{\rm ch}^{(d)}}}
=\alpha,
\eea
where $R_{\rm D}^{(k)}$ and $R_{\rm P}^{(k)}$ are given by Eqs. (\ref{tot_Rk}), respectively.

For channels satisfying (\ref{mat_iso2}), the quantities $R_{\rm D}^{(k)}$ and 
$[(d-1)/d]-R_{\rm P}^{(k)}$ tend to zero from above, as $k\to\infty$. Moreover, we obtain the 
following expression for the characteristic exponent 
\[
r_{\rm ch}^{(d)} = \ln\left[ \frac{p_{00}+(d-1)p_{01}}{p_{10}+(d-1)p_{11}}\right ] \bigg /
\ln\left[ \frac{p_{00}+(d-1)p_{01}}{p_{00}-p_{11}}\right ]. 
\]
Following \cite{RA}, Eq. (\ref{final_ineq}) can now be obtained from the condition 
for asymptotic correctability, that is $r_{\rm ch}^{(d)}>2$. However, we would like to stress that 
it is still an open problem why this particular correctability condition, which was originally 
derived for qubit-based QKD protocols, is also valid for $2d$-state protocols and 
isotropic channels.  

\section{Conclusions} 
\label{secIV}
We have discussed the error-tolerance of qudit-based QKD protocols 
using two mutually unbiased (Fourier-dual) bases. In particular, 
we focused on Gottesman-Lo-type key-distillation procedures. 
For arbitrary quantum channels subject only to the symmetry between 
the two bases used in the protocol, we derived a sufficient condition 
for secret-key distillation, thus extending known results on depolarizing 
quantum channels. 

In the case of isotropic quantum channels, we were able to analyze 
this condition further and to obtain an analytical expression for the tolerable error 
rate as a function of the dimension $d$ of the information carriers. 
Specifically, as $d\to\infty$, the tolerable error rate scales with dimension 
as $1/2-1/4\sqrt{d}$, thus approaching its upper theoretical bound, that is $1/2$. 
This asymptotic behavior is substantially different from the corresponding behavior 
in the fully symmetric $(d+1)$-basis protocols, where the tolerable error rate scales 
as $1-(3+\sqrt{5})/2d$. 

Unfortunately, for moderate values of $d$, the tolerable error rate is always well below 
its corresponding theoretical upper bound $D_{\rm th}(d)$. Hence, the development of new 
classical key-distillation protocols which will be able to bridge this gap still 
remains an interesting open problem.

\section{Acknowledgments} 
This work is supported by the EU within the IP SECOQC. K.~S.~Ranade is supported by a graduate-student 
scholarship of the Technische Universit\"at Darmstadt.

\end{document}